\documentclass[a4paper,twocolumn,11pt,unpublished]{quantumarticle}
\pdfoutput=1
\usepackage[utf8]{inputenc}
\usepackage[english]{babel}
\usepackage[T1]{fontenc}
\usepackage{hyperref}

\usepackage{tikz}
\usepackage{lipsum}

\usepackage{anysize}
\usepackage{algorithmicx,algorithm}
\usepackage{amsfonts}
\usepackage{amssymb,amsmath}
\usepackage{booktabs} % for professional tables
\usepackage{color}
\usepackage{epsf}
\usepackage{epsfig}
\usepackage{float}
\usepackage{graphicx}
\usepackage{hyperref}
\usepackage{mathrsfs}
\usepackage{microtype}
\usepackage{subfigure}
\usepackage{siunitx}
\usepackage{ulem}

\usepackage{amsmath,amsfonts,amssymb}
\usepackage{mathtools}
\usepackage{bm}
\usepackage{braket}
\usepackage{mathrsfs}
\usepackage{theorem}
\usepackage{multirow}

\newtheorem{proposition}{Proposition}

\newtheorem{theorem}[proposition]{Theorem}

\newtheorem{remark}{Remark}

\begin{document}

\title{Unleashing the Expressive Power of Pulse-Based Quantum Neural Networks}

\author{Han-Xiao Tao}
\affiliation{Center for Intelligent and Networked Systems, Department of Automation, Tsinghua University, Beijing, 100084, China}
\author{Jiaqi Hu}
\affiliation{Center for Intelligent and Networked Systems, Department of Automation, Tsinghua University, Beijing, 100084, China}
\author{Re-Bing Wu}\email{rbwu@tsinghua.edu.cn}
\affiliation{Center for Intelligent and Networked Systems, Department of Automation, Tsinghua University, Beijing, 100084, China}
%\orcid{0000-0002-2445-2701}
% \author{Christian Gogolin}
% \email{latex@quantum-journal.org}
% \homepage{http://quantum-journal.org}
% \orcid{0000-0003-0290-4698}
% \thanks{You can use the \texttt{\textbackslash{}email}, \texttt{\textbackslash{}homepage}, and \texttt{\textbackslash{}thanks} commands to add additional information for the preceding \texttt{\textbackslash{}author}. If applicable, this can also be used to indicate that a work has previously been published in conference proceedings.}
% \affiliation{Covestro Deutschland AG, Kaiser-Wilhelm-Allee 60, 51373 Leverkusen, Germany}
% \author{Marcus Huber}
% \affiliation{Institute for Quantum Optics \& Quantum Information (IQOQI), Austrian Academy of Sciences, Boltzmanngasse 3, Vienna A-1090, Austria}
% \orcid{0000-0003-1985-4623}
% \author{Cassandra Granade}
% \affiliation{Microsoft Research, Quantum Architectures and Computation Group, Redmond, WA 98052, USA}
% \author{Johannes Jakob Meyer}
% \affiliation{Dahlem Center for Complex Quantum Systems, Freie Universität Berlin, 14195 Berlin, Germany}
% \orcid{0000-0003-1533-8015}
% \author{Victor V. Albert}
% \affiliation{Institute for Quantum Information and Matter \& Walter Burke Institute for Theoretical Physics, Caltech, Pasadena, CA 91125, USA}
% \orcid{0000-0002-0335-9508}
\maketitle

\begin{abstract}
Quantum machine learning (QML) based on Noisy Intermediate-Scale Quantum (NISQ) devices hinges on the optimal utilization of limited quantum resources. While gate-based QML models are user-friendly for software engineers, their expressivity is restricted by the permissible circuit depth within a finite coherence time. In contrast, pulse-based models enable the construction of "infinitely" deep quantum neural networks within the same time, which may unleash greater expressive power for complex learning tasks. In this paper, this potential is investigated from the perspective of quantum control theory. We first indicate that the nonlinearity of pulse-based models comes from the encoding process that can be viewed as the continuous limit of data-reuploading in gate-based models. Subsequently, we prove that the pulse-based model can approximate arbitrary nonlinear functions when the underlying physical system is ensemble controllable. Under this condition, numerical simulations demonstrate the enhanced expressivity by either increasing the pulse length or the number of qubits. As anticipated, we show through numerical examples that the pulse-based model can unleash more expressive power compared to the gate-based model. These findings lay a theoretical foundation for understanding and designing expressive QML models using NISQ devices.
\end{abstract}

\section{Introduction}
Quantum computing has provable advantages over classical computing in the training, inference and sampling of machine learning processes~\cite{arute2019quantum,zhong2020quantum,zhu2022quantum,huang2022quantum}. Although the required quantum error correction is not yet available, this vision has sparked extensive researches into the implementation of quantum machine learning (QML) models on Noisy Intermediate-Scale Quantum (NISQ) devices~\cite{endo2021hybrid,callison2022hybrid}, hoping that quantum supremacy can be achieved before large-scale error-correctible quantum computers are available.

The currently adopted QML models are mostly implemented by parametrized quantum circuits, also known as quantum neural networks (QNNs)~\cite{farhi2018classification,mcclean2018barren}, whose gate parameters are trained by some external classical optimization algorithm. The state-of-art NISQ hardware has been able to support QML models that are not classically simulatable, however, only shallow circuits can be implemented due to the limited coherent time. This restricts the expressivity of the models because the number of trainable parameters is proportional to the circuit depth.

From a hardware perspective, QML models may also be parameterized by continuous-time control pulses instead of discrete-time gate parameters. Compared with gate-based models whose circuit ansatz structure may not be hardware efficient, pulse-based models is natuarally compatible with the hardware topology. This merit makes it useful for the expressivity analysis when considering the impact of the actual hardware topology. 

Owing to the proximity to the underlying physical system, pulse-based models are much easier to design and implement, and thus is friendly to hardware engineers~\cite{wu2020end,pan2023experimental,liang2022variational,melo2023pulse,ibrahim2022evaluation}. Moreover, pulse-based QML models are potentially more expressive than the gate-based models because they form "infinitely" deep QNNs within the same coherence time. 

This paper will investigate this potential via expressivity analysis of pulse-based QNN models. In the literature, the expressivity of gate-based QNNs has been discussed from different perspectives. In the view of universal quantum computation~\cite{du2020expressive, morales2020universality,biamonte2021universal,chen2021universal}, Sim \emph{et al.}~\cite{sim2019expressibility} proposed a measure for expressiveness by the uniformity of the distribution of the unitary transformations realized by QNNs over the unitary group. However, the uniformity does not necessarily imply that the underlying QML model is capable of approximating arbitrary nonlinear functions~\cite{schuld2021effect}. 

The expressivity of gated-based models can also be quantified by the complexity measures such as the pseudo-dimension~\cite{caro2020pseudo}, covering number~\cite{du2022efficient}, Rademacher complexity~\cite{bu2022statistical}, and effective dimension~\cite{abbas2021power}, which collectively indicate that the expressive power is highly dependent on the circuit depth and the number of parameters. To compare with classical NNs, it was found that~\cite{wright2020capacity}, by the measure of memory capacity, QNNs could have exponentially larger capacities than their classical counterparts.

From the perspective of kernel-based learning, the exponentially large quantum Hilbert space provides the basic feature space for embedding nonlinear functions~\cite{havlivcek2019supervised,schuld2019quantum,schuld2021supervised}. On top of it, the expressive power can be further enhanced by introducing higher nonlinearity by replicating data registers at the cost of increased circuit width or interleaving the training circuits by repeated data-uploading blocks at the cost of increased circuit depth~\cite{goto2021universal,wu2021expressivity,PerezSalinas2020datareuploading,PhysRevA.104.012405}. Fourier analysis shows that such QNN models can approximate any nonlinear functions with a spectrally rich encoding Hamiltonian as long as the numbers of qubits and circuit layers are sufficiently large~\cite{schuld2021effect,gan2022fock}.

The expressivity of pulse-based QNN models can be understood from the controllability of the underlying quantum system~\cite{lloyd1995}, which characterizes the ability of generating arbitrary unitary transformations via the controlled quantum dynamics. However, it can be proven that fully controllable pulse-based QNN models always converge to unitary $t$-designs in the long-time limit~\cite{banchi2017driven}, which may lead to barren plateaus that must be avoided in practice~\cite{larocca2022diagnosing}. 

To the authors' knowledge, the universal approximation property of pulse-based QNN models has not been explored. In this paper, we will investigate this problem from the perspective of quantum control theory, demonstrating that the pulse-based model can approximate arbitrary nonlinear functions when the underlying physical system is ensemble controllable. Numerical simulations show that the pulse-based model can unleash more expressive power from the gate-based model.

The remainder of this paper is arranged as follows. Section~\ref{Gate-based QML models} introduces the pulse-level model as the continuous limit of gate-based data-reuploading models. Section~\ref{Expressivity of QML models} introduces the ensemble controllability as a sufficient condition for the examination of expressivity of pulse-based models. In Section~\ref{Numerical experiments}, numerical experiments are performed to demonstrate the results and the effect of system dimension and the Hamiltonian structure on the model expressivity, as well as the comparison with gate-based models. Finally, concluding remarks are made in Section~\ref{Conclusion}.

\section{From gate-based to pulse-based QNN models}
\label{Gate-based QML models}
\begin{figure}[htbp]
  \centering
  {\includegraphics[width=1\columnwidth]{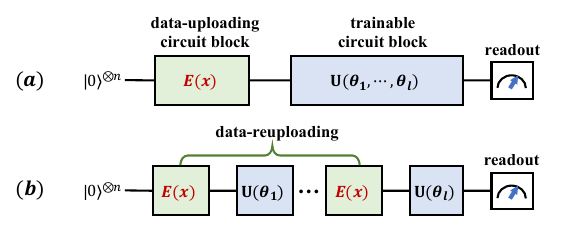}}\hfill
  \caption{Schematics of gate-based models: (a) a standard QNN model and (b) a data re-uploading QNN model.}
  \label{Fig:gate-level}
\end{figure}

Consider a standard gate-based QNN model shown in Fig.~\ref{Fig:gate-level}(a). The data-uploading circuit encodes the input data $x$ onto the quantum state of an $n$-qubit quantum register, which is then processed by a trainable circuit parameterized by $\Theta=(\theta_1,\cdots,\theta_l)$. The final learning result is yielded by the expectation-value measurement of some observable $M$. 

Suppose that the register is initialized at state $\ket{0}^{\otimes n}$. Let $U(\bm{x};\Theta)$ be the unitary transformation generated by the QNN model. The mapping from the data to the output can be described as follows: 
\begin{equation}\label{eq:mapping}
f(x;\Theta)=\bra{0}^{\otimes n}U^{\dagger}(x;\Theta)MU(x;\Theta)\ket{0}^{\otimes n}.
\end{equation}

The model is said to have universal approximation property if the mapping \eqref{eq:mapping} can approximate any complex nonlinear functions. To enhance the required nonlinearity, one can restructure the model by inserting duplicated data-encoding blocks between parameterized circuit sub-blocks, as is shown in Fig.~\ref{Fig:gate-level}(b). The resulting data-reuploading model can significantly increase the nonlinearity of $U(x;\Theta)$ at the price of increased circuit depth~\cite{PerezSalinas2020datareuploading,PhysRevA.104.012405,schuld2021effect}.

%In practice, one needs to fit the model with a given set of data samples with some proper $\bm{\theta}$ via optimization. On NISQ devices, the tuning of $\bm{\theta}$ has to be performed using classical optimization algorithms. Consequently, such variational QML algorithm is also known as hybrid quantum-classical algorithms.

\begin{figure}[htbp]
\centering
\includegraphics[width=1\columnwidth]{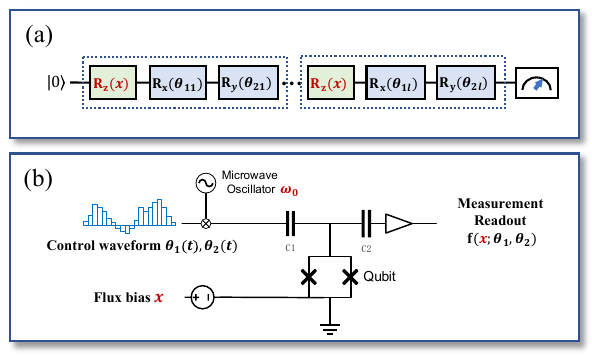}\hfill
\caption{Schematics of (a) a single-qubit gate-based data re-uploading QNN and (b) the corresponding superconducting quantum circuit implementation.}
\label{Fig:pulse-level}
\end{figure}

To see how the gate-model can be transformed to a pulse-based model, let us consider a single-qubit data-reuploading model shown in Fig.~\ref{Fig:pulse-level}(a) with $l$ encoding layers that uploads the input data $x$ through the $z$-axis rotation gate $R_{\rm z}(x)=\exp(-ix\sigma_{\rm z})$. The trainable blocks are applied by ${\rm x}$-axis rotations $R_{\rm x}(\theta_{1k})$ and ${\rm y}$-axis rotations $R_{\rm y}(\theta_{2k})$. 

Now we set the rotating angles to be infinitesimal, i.e., $x\rightarrow x\Delta t$, $\theta_{1k}\rightarrow \theta_{1k}\Delta t$ and $\theta_{2k}\rightarrow \theta_{2k}\Delta t$ with $\Delta t\rightarrow 0$. Meanwhile, we keep $n\Delta t =T$ fixed by pushing $n\rightarrow \infty$. According to the Trotter's formula
\begin{eqnarray}
  && e^{-i\theta_{2k}\Delta t\sigma_{\rm y}}e^{-i\theta_{1k}\Delta t\sigma_{\rm x}}e^{-ix\Delta t \sigma_{\rm z}}\nonumber \\
  &\approx& e^{-i(x \sigma_{\rm z}+\theta_{1k}\sigma_{\rm x}+\theta_{2k}\sigma_{\rm y})\Delta t},
\end{eqnarray}
the continuous limit leads to a pulse-based QNN model described by the following Schr\"{o}dinger equation:
\begin{small}
\begin{equation}
\label{quantum system in simple case}
\ket{\dot\psi(t;x)}=-i\left[x\sigma_{\rm z}+\theta_1(t)\sigma_{\rm x}+\theta_2(t)\sigma_{\rm y}\right]\ket{\psi(t;x)}
\end{equation}
\end{small}
that is parameterized by continuous-time control functions $\theta_1(t)$ and $\theta_2(t)$. Different from the gate-based model, the data-uploading Hamiltonian $\sigma_{\rm z}$ and the training Hamiltonians $\sigma_{\rm x}$ and $\sigma_{\rm y}$ act simultaneously (instead of alternatively) on the quantum state. 

As is illustrated in Fig.~\ref{Fig:pulse-level}, this model can be physically realized by a frequency tunable superconducting qubit dicated by a DC flux bias $x$. The transition between the qubit's logical states is manipulated by the microwave Rabi control pulse $\theta(t)=\theta_1(t)\cos\omega t+\theta_2(t)\sin\omega t$, where $\omega$ is the frequency of the local oscillator. Since the quadrature components $\theta_1(t)$ and $\theta_2(t)$ are usually generated in a piecewise-constant waveform, the unitary transformations generated by the sub-pulses can be taken as the QNN layers. Different from the gate-based models, the number of pulse-based QNN layers can be increased by shortening the sampling period $\Delta t$ without increasing the operation time, and thus we can in principle construct "infinitely" deep QNN models.  

The same procedure can be extended to the fitting of multi-variable functions using multiple qubits. In this paper, we consider the following type of controlled quantum systems:
\begin{equation}
\label{quantum system}
\ket{\dot\psi(t;\textbf{x})}=-i\left[H_0(\textbf{x})+\sum_{k=1}^p\theta_k(t)H_{k}\right]\ket{\psi(t;\textbf{x})},
\end{equation}
in which the input variable $\textbf{x}=(x_1,\cdots,x_m)$ in some range $\mathcal{X}\subseteq \mathbb{R}^m$ is uploaded through the Hamiltonian 
\[
H_0(\textbf{x})=x_1D_1+\cdots+x_mD_m
\] with $D_1,\cdots,D_m$ being the associated data-encoding Hamiltonians. Similarly, the expectation-value measurement
\begin{equation}\label{eq:measurement}
  f(\textbf{x};\Theta(t))=\langle \psi(T;\textbf{x})|M|\psi(T;\textbf{x})\rangle,
\end{equation}
is applied to approximate multi-variable functions by training the control function $\Theta(t)=(\theta_1(t),\cdots,\theta_p(t))$.

The pulse-based model is by nature hardware efficient because the Hamiltonian directly describes the hardware qubit-connectivity topology. In principle, any experimentally tunable time-dependent quantities can be applied to parameterize the pulse-based QML model. For example, in later numerical simulations, we will use the following circularly coupled $n$-qubit system
\begin{widetext}
\begin{equation}\label{eq:simulation}
H[x;\Theta(t)]=x\sum_{k=1}^n\sigma_{\rm z}^{(k)}+\sum_{k=1}^n \left[\theta_x^{(k)}(t)\sigma_{\rm x}^{(k)}
+\theta_y^{(k)}(t)\sigma_{\rm y}^{(k)}\right]+\theta_z^{(1)}(t)\sigma_{\rm z}^{(1)}\sigma_{\rm z}^{(2)}+\cdots+\theta_z^{(n)}(t)\sigma_{\rm z}^{(n)}\sigma_{\rm z}^{(1)},
\end{equation}
\end{widetext}
where $\sigma_\alpha^{(k)}=\mathbb{I}_2\otimes\cdots\otimes\sigma_\alpha\otimes\cdots\otimes \mathbb{I}_2$ with $\sigma_\alpha$ on the $k$th site, to encode a single-variable function. In addition to the local control fields $\theta_{\rm x,y}^{(k)}(t)$, the tunable qubit-qubit couplings $\theta_{\rm z}^{(k)}(t)$ can also be applied to parameterize the model.

\section{Ensemble controllability for the expressivity of pulse-based models}
\label{Expressivity of QML models}
The pulse-based model~\eqref{quantum system} is said to be expressive if, given an arbitrary function $f_0(\textbf{x})$ and an error threshold $\epsilon>0$, there always exist a proper control pulse $\Theta(t)$ such that the measurement-induced function \eqref{eq:measurement} can approximate it with the prescribed precision, i.e., $\|f(\textbf{x};\Theta(t))-f_0(\textbf{x})\|<\epsilon$.  

To see how the model can be expressive, let us first consider the fitting of a finite collection of training data samples $(\textbf{x}^{(1)},y^{(1)}),\cdots,(\textbf{x}^{(N)},y^{(N)})$. This can be treated as the manipulation of $N$ non-interacting dynamical processes:
\begin{small}
\begin{equation}\label{eq:X-equation}
\ket{\dot{\psi}(t;\textbf{x}^{(j)})} =-i\left[
H_0(\textbf{x}^{(j)})+\sum_{k=1}^p\theta_k(t)H_{k}\right]\ket{{\psi}(t;\textbf{x}^{(j})}, \nonumber
\end{equation}
\end{small}
where $j=1,\cdots,N$, under the same control function $\Theta(t)$. In the continuous limit, the data fitting (or function approximation) leads to the control of an ensemble system consisting of innumerably many non-interacting subsystems labeled by $\mathbf{x}\in\mathcal{X}$. 

To perfectly fit the dataset, there should exist a control function $\Theta(t)$ over a time period $[0,T]$ under which  $$\bra{\psi(T;\textbf{x}^{(j)})}M\ket{\psi(T;\textbf{x}^{(j)})}=y^{(j)}$$ for all $j=1,\cdots,N$. This can be done if the control is able to steer the quantum states to
\begin{eqnarray}   \ket{\psi(T;\textbf{x}^{(j)})}&=&\sqrt{\frac{y^{(j)}-\lambda_{\max}}{\lambda_{\min}-\lambda_{\max}}}\ket{\lambda_{\min}} \nonumber \\
   &&+\sqrt{\frac{y^{(j)}-\lambda_{\min}}{\lambda_{\max}-\lambda_{\min}}}\ket{\lambda_{\max}},
\end{eqnarray}
where $\ket{\lambda_{\min}}$ and $\ket{\lambda_{\max}}$ are the eigenvectors of $M$ that correspond to the smallest and largest eigenvalues.

Therefore, the dataset can be perfectly fitted when the control function is capable of simultaneously steering the $N$-tuple state $\{\ket{\psi(T;\textbf{x}^{(1)})},\cdots,\ket{\psi(T;\textbf{x}^{(N)})}\}$ to arbitrary desired final states. Similarly, to approximate continuous-variable nonlinear functions, a sufficient condition is that the ensemble state $\ket{\psi(t;\textbf{x})}$ can be steered to arbitrarily close to any desired ensemble state $\ket{\psi_{\rm f}(\textbf{x})}$ at the final time $t=T$.

The ability of driving a single quantum system to arbitrary states at some finite time is known as the controllability that has been extensively studied in the literature~\cite{albertini2003notions}. For a $d$-dimensional quantum control system $$H(t)=H_0+\sum_{k=1}^p\theta_k(t)H_{k},$$the system is controllable when the Lie algebra generated by nested commutators between the drift and control Halmiltonians $iH_0,iH_{1},\cdots,iH_{p}$ fills up $su(d)$ that consists of all traceless skew-Hermitian matrices. 
%In addition, all states can be reached within an arbitrarily small time if the Lie algebra generated by the control Halmiltonians $iH_{1},\cdots,iH_{p}$ (without the drift Hamiltonian) fills up $su(d)$.
This Lie algebra rank condition (LARC) indicates that, for the controllability examination on a curved manifold, it is sufficient to check all locally viable evolution directions generated by the control fields. 

For the ensemble system \eqref{quantum system} that involves data-uploading Hamiltonians, we need the so-called ensemble controllability~\cite{Li2006,zhang2021ensemble}. The corresponding LARC~\cite{zhang2021ensemble} requires that the Lie algebra generated by $iH_0(\textbf{x}),iH_{1},\cdots,iH_{p}$ in \eqref{quantum system} contains all $\textbf{x}$-dependent skew-Hermitian matrices $\mathcal{L}(x)$ in $su(d)$. 
In particular, for the class of functions that can be Taylor expanded in a neighborhood of $\textbf{x}=0$, the expressivity criterion can be summarized in the following theorem~\cite{zhang2021ensemble}.
\begin{theorem}\label{thm:universal approximation properties}
Let $\mathcal{X}\subseteq\mathbb{R}^m$ be a connected domain that contains $\textbf{x}=0$ as an interior point. The pulse-based model~\eqref{quantum system} can approximate any analytic functions $f: \mathcal{X}\rightarrow[\lambda_{\min},\lambda_{\max}]$ if the Lie algebra generated by $iH_0(\textbf{x}),iH_{1},\cdots,iH_{p}$ is equal to 
\begin{equation}
\mathscr{L}=span\{x_1^{r_1}\cdots x_m^{r_m},r_1,\cdots,r_m\in \mathbb{N}\}\otimes su(d).  \nonumber
\end{equation} 
\end{theorem}

Since the condition of Theorem \ref{thm:universal approximation properties} is often uneasy to examine for multivariate cases, we present the following easy-to-check expressivity condition.
\begin{theorem}\label{thm:universal approximation properties_2}
The pulse-based model~\eqref{quantum system} can approximate any function $f: \mathcal{X}\rightarrow[\lambda_{\min},\lambda_{\max}]$ if the Lie algebra $\{iH_{1},\cdots,iH_{p}\}_{LA}=su(d)$.
\end{theorem}

\noindent {\bf Proof:} According to Theorem~\ref{thm:universal approximation properties}, it is sufficient to prove that  $ix_1^{r_1}\cdots x_m^{r_m}H$ is contained in the Lie algebra generated by $H_0(\textbf{x}),H_{1},\cdots,H_{p}$ for any $(r_1,\cdots,r_m)$ and any $H\in su(d)$. This can be inductively proved as follows. 

First, the condition that $\{H_1,\cdots,H_p\}_{LA}=su(d)$ implies that the above statement holds for $r_1=\cdots=r_m=0$. Now suppose that the statement holds for some $r_1,\cdots,r_m$, and we prove its correctness for $r_1,\cdots,r_j+1,\cdots,r_m$ with $j=1,\cdots,m$. 

Let $\lambda_1,\cdots,\lambda_d$ be the eigenvalues of the data-uploading Hamiltonian $D_j$ associated with $x_j$, and $\ket{\lambda_1},\cdots,\ket{\lambda_d}$ are the corresponding eigenvectors. Under this basis, the Lie algebra $su(d)$ can be expanded by the following group of skew-Hermitian operators:
\begin{eqnarray}
X_{\alpha\beta}&=&i(\ket{\lambda_\alpha}\bra{\lambda_\beta}+\ket{\lambda_\beta}\bra{\lambda_\alpha}),\nonumber\\
Y_{\alpha\beta}&=& \ket{\lambda_\alpha}\bra{\lambda_\beta}-\ket{\lambda_\beta}\bra{\lambda_\alpha},\nonumber \\
Z_{\alpha\beta}&=&i(\ket{\lambda_\alpha}\bra{\lambda_\alpha}-\ket{\lambda_\beta}\bra{\lambda_\beta}),\nonumber
\end{eqnarray}
where $1\leq \alpha <\beta\leq d$. 

When $D_j$ has a non-degenerate spectrum, it is easy to verify the following commutation relations:
\begin{widetext}
\begin{eqnarray}
~[ix_jD_j,ix_1^{r_1}\cdots x_j^{r_j}\cdots x_m^{r_m}X_{\alpha\beta}]&=&ix_1^{r_1}\cdots x_j^{r_j+1}\cdots x_m^{r_m}Y_{\alpha\beta}, \\
~[ix_jD_j,ix_1^{r_1}\cdots x_j^{r_j}\cdots x_m^{r_m}Y_{\alpha\beta}]&=&ix_1^{r_1}\cdots x_j^{r_j+1}\cdots x_m^{r_m}X_{\alpha\beta}, \\
~[ix_1^{r_1}\cdots x_j^{r_j+1}\cdots x_m^{r_m}X_{\alpha\beta},iY_{\alpha\beta}]&=&x_1^{r_1}\cdots x_j^{r_j+1}\cdots x_m^{r_m}Z_{\alpha\beta},
\end{eqnarray}
\end{widetext}
for all $1\leq \alpha<\beta\leq d$, which implies that all terms $x_1^{r_1}\cdots x_j^{r_j+1}\cdots x_m^{r_m}H$ with arbitrary $H\in su(d)$ can be generated, and this holds for all $j=1,\cdots,m$.

When $D_j$ has degenerate eigenvalues, say $\lambda_\alpha=\lambda_\beta$, we can pick some $\lambda_\gamma$ that is different from them, so that $ix_1^{r_1}\cdots x_j^{r_j+1}\cdots x_m^{r_m}Z_{\alpha\gamma}$ and $ix_1^{r_1}\cdots x_j^{r_j+1}\cdots x_m^{r_m}Z_\beta{\gamma}$ can be generated as above. These two terms can be used to generate $ix_1^{r_1}\cdots x_j^{r_j+1}\cdots x_m^{r_m}Z_{\alpha\beta}$ from the fact that $Z_{\alpha\gamma}-Z_{\beta\gamma}=Z_{\alpha\beta}$. Further, the commutators of $ix_1^{r_1}\cdots x_j^{r_j+1}\cdots x_m^{r_m}Z_{\alpha\beta}$ with $iX_{\alpha\beta}$ and $iY_{\alpha\beta}$ produce $ix_1^{r_1}\cdots x_j^{r_j+1}\cdots x_m^{r_m}Y_{\alpha\beta}$ and $ix_1^{r_1}\cdots x_j^{r_j+1}\cdots x_m^{r_m}X_{\alpha\beta}$, respectively. This ends the proof.

\begin{remark}
It should be noted that the ensemble controllability is a sufficient but not necessary condition for the expressivity of pulse-based QNN models. This condition guarantees that the model is expressive no matter which measurement observable $M$ is chosen and no matter what initial state is selected. When the system is not ensemble controllable, the model may still be expressive, but this depends on the choice of the proper observable and the initial state.   
\end{remark}

Now let us take the single-qubit model~\eqref{quantum system in simple case} as an example. Simple calculations show that the generated Lie algebra
$$\{x\sigma_{\rm z},\sigma_{\rm x},\sigma_{\rm y}\}_{LA}=\{x^{k},k\in \mathbb{N}\}\otimes su(2)$$
fulfils the Lie algebraic condition in Theorem~\ref{thm:universal approximation properties} and Theorem~\ref{thm:universal approximation properties_2}. As will be numerically simulated in the next section, this model is able to approximate arbitrary nonlinear functions.

However, if the control on $\sigma_{\rm y}$ is turned off, the underlying Lie algebra
$$
\{x\sigma_{\rm z},\sigma_{\rm x}\}_{LA}=\{x^{2k}\sigma_{\rm x},x^{2k+1}\sigma_{\rm y},x^{2k+1}\sigma_{\rm z},k\in \mathbb{N}\}$$
does not contain all monomials of $\sigma_{\rm x},\sigma_{\rm y},\sigma_{\rm z}$, and hence the system is not ensemble controllable. As is indicated in \cite{yu2022power}, this model is not fully expressive because only even (odd) functions of $x$ can be approximated when $M=\sigma_{\rm y}$ ($M=\sigma_{\rm x}$).

\section{Numerical experiments}
\label{Numerical experiments}
In this section, we will perform numerical experiments to examine the expressivity of ensemble controllable systems. 

For simplicity, the observable $M$ is always chosen such that $\lambda_{\min}=-1$ and $\lambda_{\max}=1$. The domain of functions to be fitted is assumed to be a hypercube $\mathcal{X}=[-R,R]^m\in\mathbb{R}^m$. Under the following rescaling transformation 
\begin{equation}
    \textbf{x}=R\bar{\textbf{x}},\quad \bar{t}=Rt,\quad \bar{\Theta}(t)=R^{-1}\Theta(t),
\end{equation}
under which the controlled Schr\"{o}dinger equation is formally invariant:   
\begin{equation}
  \ket{\dot{\psi}(\bar t;\bar{\textbf{x}})}=-i\left[H_0(\bar{\textbf{x}})+\sum_{k=1}^p\bar{{\theta}}_k(t)H_k\right]\ket{\psi(t;\bar{\textbf{x}})}.
 \end{equation}
Therefore, we can always assume that $\mathcal{X}=[-1,1]^m$ without loss of generality. 

In our numerical experiments, we apply the Adam optimizer~\cite{kingma2014adam} to train pulse-based models by minimizing the mean squared error (MSE) 
\begin{equation}
L[\Theta(t)]=N^{-1}\sum_{k=1}^N\left\|f(\textbf{x}^{(k)};\Theta(t))-y^{(k)}\right\|^2
\end{equation}
over the training dataset $\{(\textbf{x}^{(k)},y^{(k)})\}$ sampled from the target function to be fitted. 

\subsection{The approximation of nonlinear functions}
\begin{figure}[htbp]
\centering
\includegraphics[width=1\columnwidth]{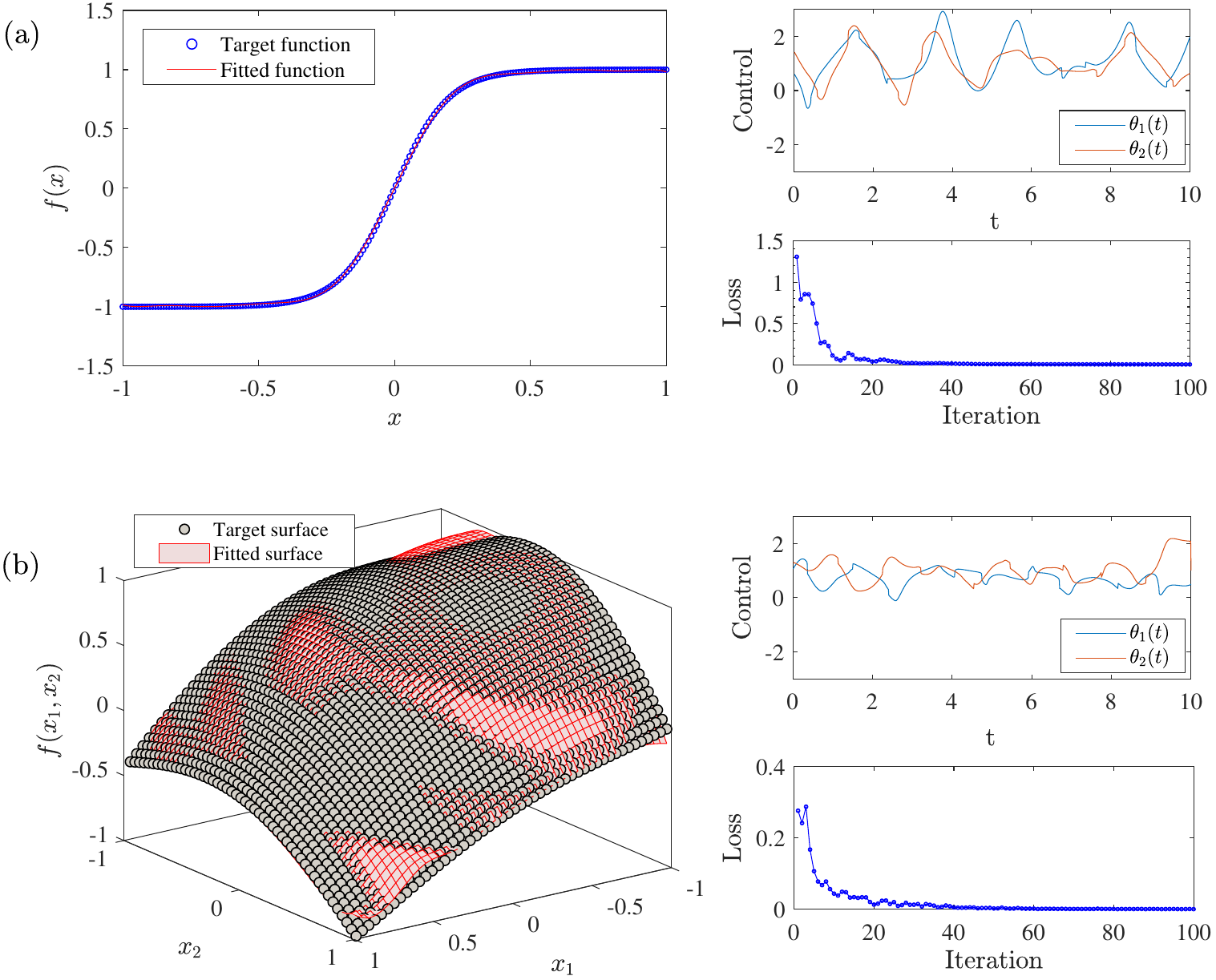}\hfill
\caption{The approximation results of (a) univariate function and (b) bivariate function by single-qubit pulse-based models. On the left are the fitted curve and surface, and on the right are the resulting control fields and the training curves.}
\label{Fig:univariate function and bivariate functions}
\end{figure}

We start from the single-qubit model \eqref{quantum system in simple case}. The target function is chosen as the sigmoid-like function
\begin{eqnarray}
\label{sigmoid}
  f_{1}(x) &=& \frac{1-e^{-10x}}{1+e^{-10x}},
\end{eqnarray}
from which 200 points are evenly sampled from $x\in [-1,1]$ to train the model. The measurement observable is chosen as $M=\sigma_{\rm z}$, while the time duration and the number of subpulses are chosen to be $T=10$ and $K=1000$. As is shown in Fig.~\ref{Fig:univariate function and bivariate functions}(a), the trained model almost perfectly fit the data points after 100 rounds of iteration. 

The single-qubit model may also approximate more complicated functions. For illustration, we use the following system 
\begin{equation}\label{eq:example3}
  H(t) = x_1\sigma_{\rm x}+x_2\sigma_{\rm y}+\theta_1(t)\sigma_{\rm x}+\theta_2(t)\sigma_{\rm z}
\end{equation}
to approximate a bivariate function 
\begin{small}
\begin{equation}
  f(x_1,x_2)=(x_1^2+x_2-1.5\pi)^2+(x_1+x_2^2-\pi)^2
\end{equation}
\end{small}%
that was simulated in Ref.~\cite{yu2022power} with gate-based models. In the simulation, $50\times 50=2500$ data points are evenly sampled over the domain $\mathcal{X}=[-1,1]\times[-1,1]$. The measurement observable and hyperparameters $T$ and $K$ are kept the same. The simulation results shown in Fig.~\ref{Fig:univariate function and bivariate functions}(b) indicates that the trained pulse-based model can well fit the surface of the target bivariate function after 100 rounds of iteration, which is much better than that of gate-based circuit models~\cite{yu2022power}.  

%\begin{figure}[htbp]
%\centering
%\includegraphics[width=1\columnwidth]{./images/1q-odd}\hfill
%\caption{The fitting results of a sigmoid function by a single-qubit model with insufficient expressivity.}
%\label{Fig:odd function}
%\end{figure}

\subsection{The affection of pulse duration}
\begin{figure}
    \centering
    \includegraphics[width=1\linewidth]{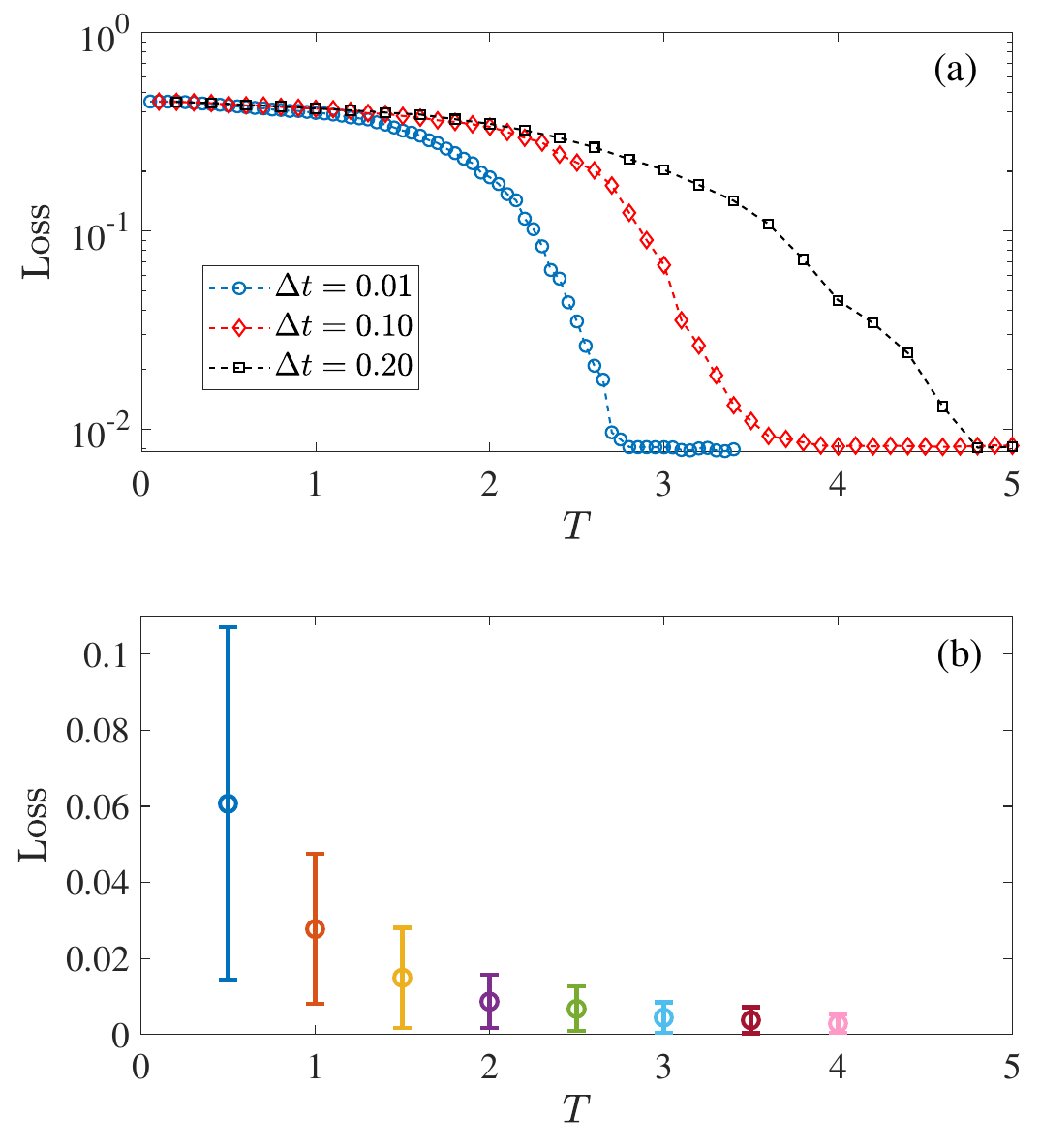}
    \caption{The training loss versus the pulse duration time $T$: (a) fitting the univariate function~\eqref{function_2} with different sampling periods; (b) the statistics of training losses over a class of sampled polynomial functions.}
    \label{fig:LvsT}
\end{figure}

In the training of pulse-based models, the pulse duration time $T$ and the number $K$ of discretized subpulses are crucial hyperparameters to be pre-selected. Figure \ref{fig:LvsT} displays how the training loss varies with the pulse duration $T$ for the fitting of univariate function
\begin{small}
\begin{eqnarray}
\label{function_2}
  f_{2}(x) &=& 10x^2-14x^4-3x^6+7x^8-\cos x.
\end{eqnarray}
\end{small}%
using the single-qubit model \eqref{quantum system in simple case}. It can be seen that the loss monotonically decreases with $T$ when the sampling period $\Delta t=T/K$ is fixed, but the loss stops decreasing when $T$ is above some threshold value. This implies that "deeper" QNNs are more expressive, and the expressivity saturates when the QNN is sufficiently deep. 

On the other hand, one can construct deeper QNNs by finer discretization without increasing $T$. The comparison between different discretization schemes show that the overall expressivity increases when $\Delta t=T/K$ gets smaller, but the advantage vanishes when $\Delta t$ is sufficiently small.

%The curves plotted in Fig.~\ref{fig:LvsT}(a) provide a good measure for the degree of expressivity. A pulse-based model is more expressive if the curve has a steeper falling slope for all target functions. The threshold time for reaching some threshold error indicates the shortest pulse length that can be trained.  
%A mTo fit a function by an ensemble controllable system with some error threshold $\epsilon$, there is always a minimal time $T_{\min}$ below which the training loss cannot be lower than $\epsilon$ no matter how small $\Delta t$ is. Also, for a fixed $T>T_{\min}$, there is a minimal $K_{\min}$ below which the desired precision is not reachable. The estimation of these threshold parameters are practically important for designing more trainable pulse-based QNN models.

To further test the expressivity of the model, we train the model for $100$ randomly selected polynomial functions in the following form:
\begin{equation}
  f(x)=\sum_{j=1}^8a_jx^j, 
\end{equation}
where the coefficients are uniformly sampled in $[-30,30]$. Figure~\ref{fig:LvsT}(b) shows that the increase of pulse duration effectively reduces the average training loss and meanwhile narrows down the distribution, which means that the ability of the model to approximate arbitrary functions increases with $T$. This clearly indicates the universal approximation property of the model when the pulse duration is sufficiently long.

\begin{figure}[htbp]
\centering
\includegraphics[width=1\columnwidth]{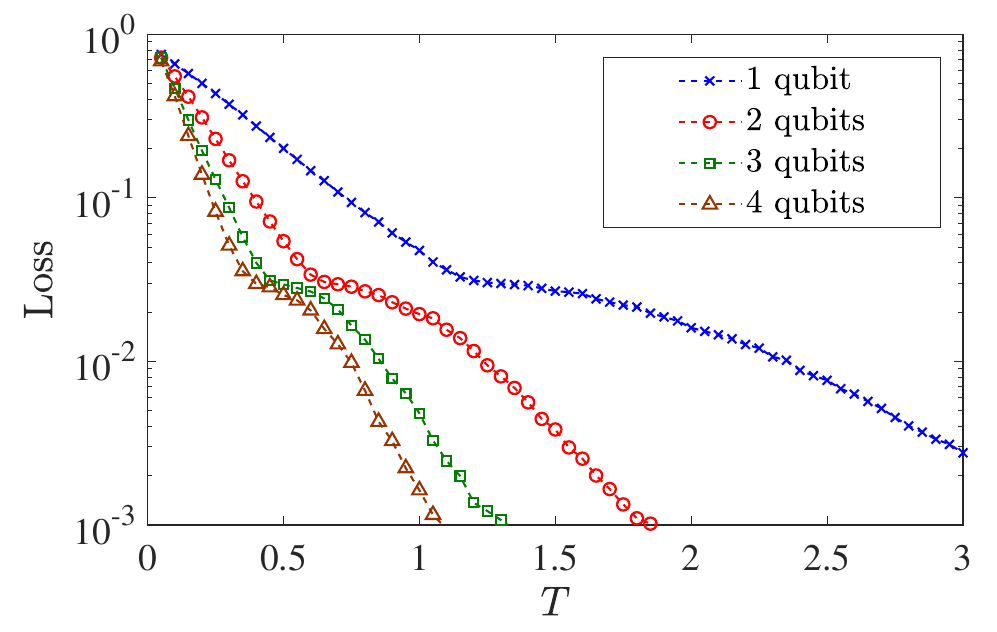}\hfill
\caption{The dependence of training loss on the depth and width of the pulse-based models with circularly coupled qubits.}
\label{Fig:T-Loss-Qnum}
\end{figure}

In addition to the duration time $T$ that characterizes the "depth" of the pulse-based QNN models, the expressive power is also dependent on the number $n$ of qubits that corresponds to its width. For illustration, we fit the sigmoid-like function \eqref{sigmoid} using the circularly coupled multi-qubit model~\eqref{eq:simulation} with up to four qubits. It can be verified that this class of models is ensemble controllable as it satisfies the criterion proposed by Theorem~\ref{thm:universal approximation properties_2}. Their training-loss curves are plotted in Fig.~\ref{Fig:T-Loss-Qnum} versus the pulse duration $T$, where the sampling period is fixed as $\Delta t=0.01$. 

As is predicted by Theorem~\ref{thm:universal approximation properties_2}, all models can precisely approximate the sigmoid function owing to the ensemble controllability. The utilization of more qubits can shorten the required pulse duration, implying that larger models have higher expressivity than small models. However, less improvement is gained when $n$ grows, e.g., the gap between $n=3$ and $n=4$ is remarkably smaller than that between $n=1$ and $n=2$. This is reasonable because the function to be approximated is not complicated so that the increase of model expressivity will not bring additional benefits. Large-size models are expected to be more advantageous when approximating complex functions. 

\subsection{Unleashing the expressive power from gate-based models}
Intuitively, pulse-based models are more expressive than gate-based models deployed on the same hardware, because pulse-based models are more hardware efficient and can be designed deeper within the same coherence time. To examine this intuition, we train single-qubit gate-based models shown in Fig.~\ref{Fig:pulse-level}(a) and estimate their overall gate operation time based on the same physical parameters used in the pulse-based model, so that their expressive power can be compared on an equal footing.

We first train gate-based models with 5, 10 and 15 blocks, which contains 10, 20 and 30 trainable parameters, respectively. The achieved training losses are $1.6\times10^{-2}$, $1.1\times10^{-3}$ and $2.9\times10^{-4}$, respectively. Then, we train pulse-based models with identical number of parameters (i.e., $K=5$, $K=10$ and $K=15$, respectively) and gradually increase $\Delta t$ until the same training performance is achieved.

From the trained pulse-based models, the peak amplitude of the control pulses are observed to be no higher than $\theta_{\max}=2\pi\times50$MHz. Therefore, the operation time for a rotation $R_{\rm x}(\theta_{1k})$ is no smaller than $t_{1k}=\theta_{1k}/\theta_{\max}$. The gate operation time for a ${\rm y}$-rotation is calculated in the same way. Finally, we omit the operation time for $z$-axis rotations, and calculate the following lower bound
\begin{equation}
    T_{\rm G}=\theta_{\max}^{-1}\sum_{k=1}^n\left(\theta_{1k}+\theta_{2k}\right)
\end{equation}
as the estimation for the operation time of an $n$-block circuit.

\begin{figure}
    \centering
    \includegraphics[width=1\linewidth]{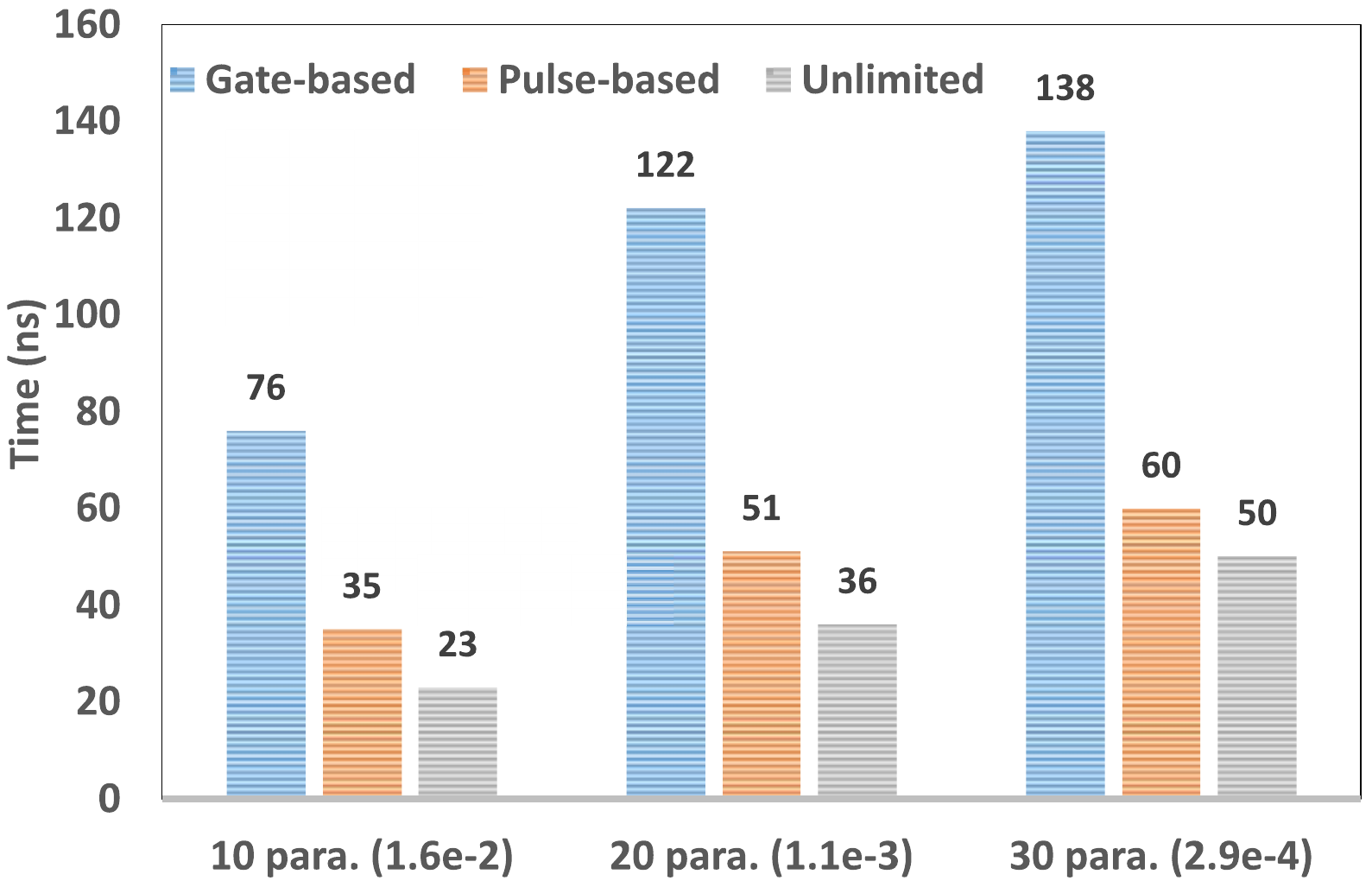}
    \caption{Comparison of the operation time with gate-based and pulse-based models that achieve the same training performances.}
    \label{fig:GvsP}
\end{figure}

The comparison of the three pairs of gate-based and pulse-based models is shown in Fig.~\ref{fig:GvsP}. Using the same number of parameters, the pulse-based models require remarkably shorter time ($<50\%$) than the corresponding lower bound of the gate-based model, showing that the inference can be made more than twice faster. 

To explore the maximal expressive power that can be unleashed, we further decrease the pulse time duration and allow unlimited number of parameters while keeping the same training performance. The resulting reduction of pulse duration time shows that there is still room for faster inference.

\section{Concluding remarks}\label{Conclusion}
To summarize, we have explored the expressive power of pulse-based QML models from a control system point of view. It is shown that the driving control pulse plays the role of data-reuploading in the continuous-time limit, and thus contributes significant nonlinearity required for the learning. We prove that pulse-based models can approximate arbitrary nonlinear functions when the underlying quantum control system is ensemble controllable. Simulations show that the expressive power increases with the model depth and width, and the pulse-based model can release more expressive power from the gate-based models. 

It should be noted that the ensemble controllability raised for expressivity is a rather strong condition. In fact, the controllability will lead to notorious barren plateaus~\cite{banchi2017driven,larocca2022diagnosing} that make the training almost impossible for large-scale models. Therefore, efficient gate-pulse transpilations~\cite{gokhale2020optimized, earnest2021pulse,ibrahim2022evaluation,melo2023pulse,niu2022effects} and co-design~\cite{liang2023hybrid} are required to achieve a good balance between the expressivity and the trainability.  

From the perspective of controllability, the model should be designed to be uncontrollable to balance the expressivity and the trainability, under which circumstances the initial states and measurements must be carefully chosen. We have found examples that are expressive but ensemble uncontrollable. How to systematically discover such learning models is an interesting open problem. The control system analysis framework also makes it possible to compare the model expressivity between hardwares with different qubit connectivity topologies. This problem will also be explored in our future studies.

% \acknowledgements
% This work is supported by the National Key R$\&$D Program of China (Grants No.~2017YFA0304304 and No.~2018YFA0306703), NSFC (Grants No.~61833010 and No.~61773232), the Key-Area R$\&$D Program of GuangDong Province (Grant No. 2018B030326001) and a grant from the Institute for Guo Qiang, Tsinghua University. Invaluable discussions with Prof. Changshui Zhang are greatly appreciated.

\end{document}